\documentclass[aps,twocolumn,10pt,groupedaddress]{revtex4-1} 
\usepackage{graphicx}  
\usepackage{subfigure}
\usepackage{dcolumn}   
\usepackage{bm}        
\usepackage{amssymb,amsmath}   

\begin{document}

\title{Providing 10$^{-16}$ short-term stability of a 1.5~$\mathbf{\mu}$m laser to optical clocks}



\author{C. Hagemann}
\altaffiliation{member of QUEST Institute for Experimental Quantum \mbox{Metrology}, Bundesallee 100, 38116 Braunschweig, Germany}
\author{C. Grebing}
\author{T. Kessler}
\altaffiliation{member of QUEST Institute for Experimental Quantum \mbox{Metrology}, Bundesallee 100, 38116 Braunschweig, Germany}
\author{St. Falke}
\author{C. Lisdat}
\author{H. Schnatz}
\author{F. Riehle}
\author{U. Sterr}
\affiliation{Physikalisch-Technische Bundesanstalt, Bundesallee 100, 38116 Braunschweig, Germany}



             
\begin{abstract}
We report on transferring \mbox{10$^{-16}$-level} fractional frequency stability of a ``master laser'' operated at 1.5~$\mathbf{\mu}$m to a ``slave laser'' operated at 698~nm, using a femtosecond fiber comb as transfer oscillator. With the 698~nm laser, the $^1$S$_0 - ^3$P$_0$ clock transition of $^{87}$Sr was resolved to a Fourier-limited line width of~1.5~Hz (before:~10~Hz). Potential noise sources contributed by the frequency comb are discussed in detail.
\end{abstract}

\maketitle  

\section{Introduction}
\label{sec:introduction}

With fractional inaccuracies of $8.6\times10^{-18}$ \cite{cho10} and instabilities of $\sigma_y=4 ... 5\times 10^{-16}/\sqrt{\tau/ {\rm s}}$ \cite{jia11} optical atomic clocks are the most precise instruments for measurements of time and frequency. Typically, a highly short-term stable interrogation laser is stabilized to a long-term stable optical transition of an atomic reference. While the stability of single-ion clocks is generally limited by quantum projection noise \cite{ita93}, optical lattice clocks with a large number of neutral atoms suffer from insufficient performance of the interrogation laser to exploit their full potential \cite{tak11,san98,que03}. 

State-of-the-art clock lasers are often stabilized to high-finesse optical cavities with spacers made of Ultra-Low Expansion (ULE) glass. Minimizing the length-sensitivity to environmental perturbations, fractional frequency stabilities of a few times 10$^{-16}$ at a second \cite{jia11,you99} have been demonstrated, but thermal noise resulting from Brownian motion inside the high-reflection coatings hampers further improvements massively \cite{num04,not06}. Hence, different approaches such as stabilizing to cryogenic single-crystal cavities \cite{kes12a}, spectral holes in prepared doped crystals \cite{str00,tho11a} or whispering-gallery-mode resonators \cite{mat07,sav07a} increasingly attract attention to lower this thermal noise.

As many of these approaches only work at specific wavelengths, transfer of an outstanding stability to the wavelength of the optical clock is needed. Utilizing a broadband optical frequency comb to bridge the spectral gap between a highly stable ``master laser'' and a less stable ``slave laser'', transfers of instabilities in the range of  10$^{-15}$ have been demonstrated \cite{gro08,leg09,yam12}. In extension, this approach allows to slave several clock lasers to one master laser simultaneously.

However, with lasers approaching sub-10$^{-16}$ instabilities, it becomes important to investigate if the transferred stability is degraded by additional noise from the transfer oscillator. In this paper this question will be addressed, utilizing a commercial femtosecond fiber comb to accomplish an octave-spanning transfer of 10$^{-16}$ instabilities to a less stable slave laser. In Section~\ref{sec:master}, we describe the master laser system providing such ultra-high stabilities. In Section~\ref{sec:transfer}, we present results of a series of noise measurements on the fiber comb to evaluate the performance of the transfer. On the example of the $^{87}$Sr optical lattice clock at PTB, the increased stability of the slave laser is demonstrated in Section~\ref{sec:sr87}.

\section{Ultra-stable master laser system}
\label{sec:master}

\begin{figure}[!b]
\centering
\includegraphics[width=\columnwidth]{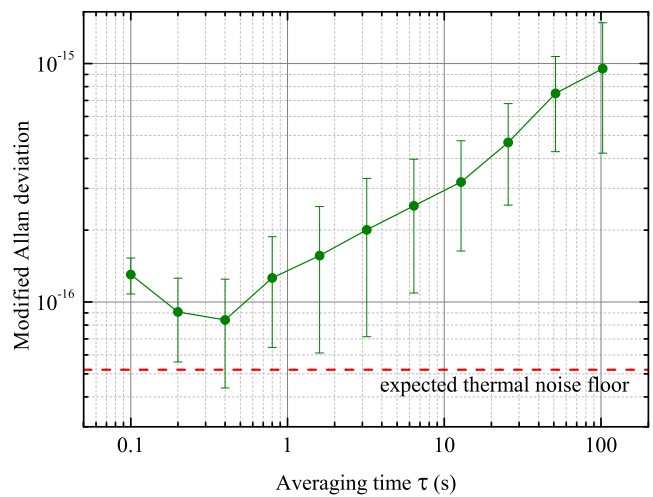}
\caption{Modified Allan deviation of the silicon cavity laser system derived from a three-cornered hat comparison with two conventional laser systems (data from Ref.~\cite{kes12a}).}
\label{fig:master}	
\end{figure}

The master laser system comprises a 1.5~$\mu$m fiber laser stabilized to an optical cavity machined from single-crystal silicon. The cavity is operated at the zero-point of its thermal expansion at a temperature of 124~K. The low operating temperature and the superior thermo-mechanical properties of silicon set the thermal noise limit \cite{num04,kes12} for the fractional frequency stability of the laser to a level of $\sigma_y\approx~6\times10^{-17}$. This value is lower by a factor of three compared to state-of-the-art resonators operated at room temperature \cite{jia11}. In a three-cornered hat comparison with two conventional laser systems the silicon cavity stabilized laser has demonstrated short-term stabilities of mod~$\sigma_y=1\ldots3\times10^{-16}$ between $0.1\ldots10$~seconds \cite{kes12a} (see Fig.~\ref{fig:master}).

Because of the absorption of the material laser transmission is restricted to the infrared between 1 and 6~$\mu$m. We chose an operation wavelength of 1.5~$\mu$m, as the laser technology at this wavelength is rather mature and the stable light can be disseminated, e.g. to the femtosecond frequency comb, at low cost using standard telecom fibers \cite{gro07}.

\section{Performance of transfer oscillator}
\label{sec:transfer}

The stability transfer from the infrared master laser to the spectral ranges of optical clock transitions is accomplished using a multi-port Er:fiber femtosecond frequency comb. We exploited the correlations of the comb offset frequency $f_{\rm{CE}}(t)$ and repetition rate $f_{\rm{rep}}(t)$ with the fluctuations of the frequency $\nu_n(t)$ of the n-th optical comb mode
\begin{equation}
	\nu_n(t)=n\cdot f_{\rm{Rep}}(t)+f_{\rm{CE}}(t)
	\label{eq:comb}
\end{equation}
which is forming the base of the transfer oscillator concept of Telle et al. \cite{tel02b}. This technique provides a transfer bandwidth in the Megahertz range as it only relies on fast radio frequency (rf) tracking electronics and no bandwidth restrictions caused by a limited response time of the mode-locked laser itself occur.

The requirements to the transfer are set by the performance of the master laser, which supports instabilities of mod~$\sigma_y=1\ldots2\times10^{-16}$ for averaging times between $0.1\ldots3$~s (see Section~\ref{sec:master}). Moreover, advanced clock comparison experiments reach residual instabilities of $\sigma_y\approx1\times10^{-17}$ \cite{ros08,fal11,hun12} for long averaging times. These numbers trigger the question whether a multi-port frequency comb is capable of bridging vast spectral gaps without adding noise to the comparison of the spectrally separated oscillators.

Equation (\ref{eq:comb}), which is the very base of the transfer concept, is valid at the output coupler of the mode-locked fiber comb. However, before the light is superimposed with the connected continuous-wave (cw) oscillators usually the light is amplified and its frequency is converted to the region of the cw oscillator. These subsequent processing steps might introduce an additional noise term $\delta \nu(n,t)$ to the right-hand side of equation (\ref{eq:comb}) \cite{new07} and thus corrupt the transfer concept.
A single broadened fiber oscillator already proved its capability of transferring stabilities at levels below $10^{-16}$ for averaging times $>1$~s \cite{swa06,cod07}. However, in commercial fiber comb systems the oscillator output is amplified and spectrally shifted or broadened to the needed regions in multiple branches. Differential noise added in these branches would compromise the result of an inter-branch frequency comparison \cite{new07,nak10}.
Therefore, prior to the stability transfer experiment itself, we have investigated and characterized the noise that is potentially added in conversion branches. In other words, we check to which level equation (\ref{eq:comb}) is valid when using a multi-port frequency comb.

\begin{figure}[!t]
	\centering
		\includegraphics[width=\columnwidth]{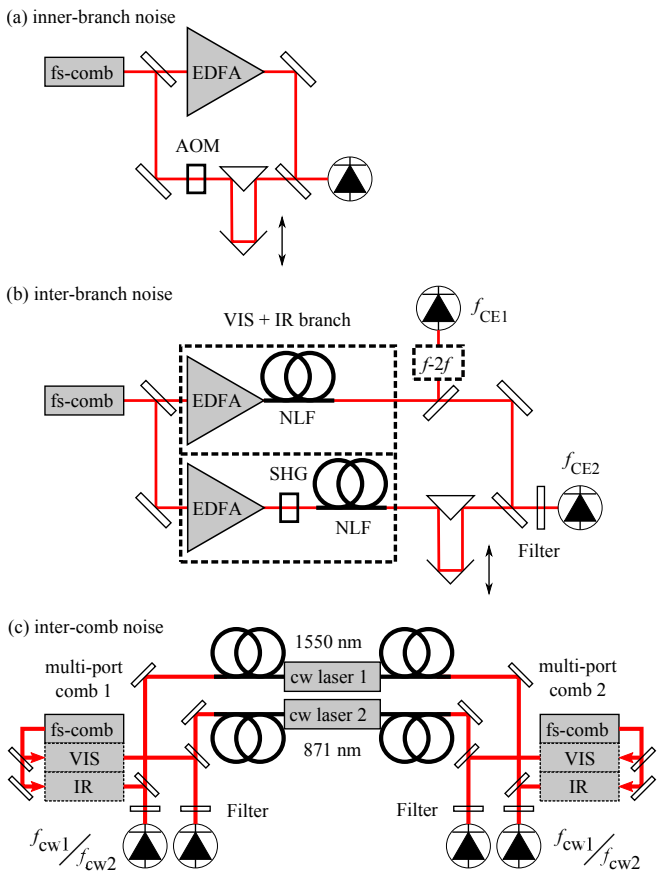}
	\caption{Measurement setups for determining and identifying the noise generated in fiber comb conversion branches. (a): Interferometer for measuring the noise added in an EDFA (inner-branch noise). (b): Inter-branch noise setup. SHG - second harmonic generation, NLF - nonlinear fiber, Filter - monochromator with a filter bandwidth of about 0.1~nm. (c): Assembly to measure a frequency ratio of two cw lasers with two independent frequency combs (inter-comb noise).}
	\label{fig:combSetup}
\end{figure}

The potential noise sources giving rise to excess noise $\delta \nu(n,t)$ were discussed in detail by Newbury et al. \cite{new07}: amplified spontaneous emission (ASE) during the amplification process, detection/technical noise, excess noise in the non-linear frequency conversion and environmental noise acting on the setup stemming from thermal or acoustic influences.
Our investigation is divided into three measurements, increasing the number of noise sources stepwise. All three arrangements are depicted in Fig.~\ref{fig:combSetup}.

$Inner$-$branch$ $noise$ (a): This setup addresses the noise generated in an erbium-doped fiber amplifier (EDFA). We fed the oscillator output into an interferometer with an EDFA in the test arm and an acousto-optic modulator (AOM) in the reference arm to allow heterodyne detection in order to trace back ASE and environmental noise contributions during the amplification process. In general, we minimized the spurious interferometer noise by building as compact as possible and passively shielded setups. The beat signal was detected with a signal-to-noise ratio (SNR) of about 50~dB (resolution bandwidth (RBW): 100~kHz).

$Inter$-$branch$ $noise$ (b): Going one step further, we added the frequency conversion step as potential excess noise source by comparing two branches that generate spectrally overlapping outputs. In detail, this was achieved by measuring the comb offset frequency $f_{\rm{CE}1}$ at about 1.1~$\mu$m in an octave-spanning IR branch ($\approx~1\ldots2.2$~$\mu$m) utilizing an $f$-$2f$ interferometer \cite{tel99}. The unused residual light is superimposed with the output of a frequency-doubled octave wide visible (VIS) branch ranging from $\approx~0.5\ldots1$ $\mu$m. The overlapping region at around 1 $\mu$m was filtered out and sent to a photo diode to again measure the comb offset frequency ($f_{\rm{CE}2}$). In each case the comb offset frequency was measured with a SNR of about 30~dB (RBW~=~100~kHz). The difference between both offset frequencies $f_{\rm{CE}1}-f_{\rm{CE}2}$ reveals the differential inter-branch noise.

$Inter$-$comb$ $noise$ (c): Finally, we built an almost fully symmetric setup sending two spectrally separated cw lasers to two completely independent multi-port frequency comb systems located in different buildings. The difference in the measured frequency ratios $f_{\rm{cw1}}/f_{\rm{cw2}}$ contains all the noise sources that can compromise such frequency comparisons. The beat signals were detected with a SNR of about 25~dB (RBW~=~100~kHz).

\begin{figure}[!t]
	\centering
		\includegraphics[width=\columnwidth]{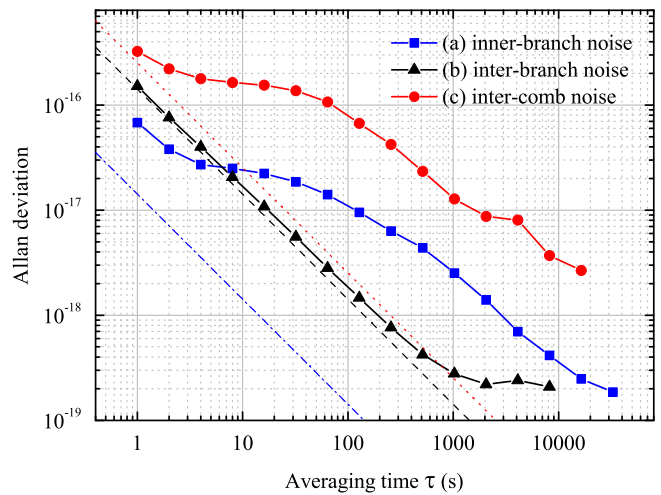}
	\caption{Fractional frequency instability in terms of Allan deviation of the different noise measurements. The straight lines indicate the detection limited instability of each measurement (assuming white phase noise, inner-branch: dashed-dotted blue, inter-branch: dashed black, inter-comb: dotted red).}
	\label{fig:combLimits}
\end{figure}

Each beat note was tracked with about $500$~kHz bandwidth using a fast tracking oscillator due to the comb line width of about 100~kHz in the optical region. The instability of each measurement was evaluated in terms of the Allan deviation and the results are depicted in Figure \ref{fig:combLimits}. 

The short-term stabilities of the all measurements are close the requirements given by the best available optical oscillators (see Section \ref{sec:master}). The observed signal-to-noise ratios of the beat signals in the inter-branch and the inter-comb measurement give on upper limit to white phase noise of $-80$~dBc/Hz and $-75$~dBc/Hz, respectively. These values convert to Allan deviations of $1\ldots3\times10^{-16}/(\tau/{\rm s})$ and are in agreement with the measured instabilities, proving the short-term performances of the inter-branch and the inter-comb measurement are limited by white phase noise. The typical detected power of a single comb mode is about $1$~nW yielding a shot-noise limited white phase noise floor $-97$~dBc/Hz, which would support instabilities in the order of $2\times10^{-17}/(\tau/{\rm s})$. The difference reveals the presence of excess phase noise generated during the non-linear broadening/shifting as verified for Ti:sapphire combs \cite{new07}~(and references therein). In the inner-branch measurement most of the comb modes contribute to the signal and excess noise resulting from spectral broadening is absent. The SNR ratio is distorted only by dispersion in the EDFA yielding a white phase noise level of below $-100$~dBc/Hz, which converts to Allan deviations of $1\ldots2\times10^{-16}/(\tau/{\rm s})$ well below the observed instability. Hence, the amplification process adds considerable noise contributions that originate from ASE and environmental noise acting on the amplification fiber.

At intermediate observation times between $1\ldots100$ seconds the inner-branch and inter-comb noise measurement show pronounced instability plateaus. This signature is known to stem from environmental perturbations acting on fiber links \cite{nak10,gro09}. A single EDFA comprises about 15~m of fiber and extra care is taken to isolate the fiber against environmental distortions. This explains the impressively low level of the plateau at low $10^{-17}$ in the inner-branch measurement. 
The fiber noise plateau in the inter-comb experiment at $1\ldots2\times10^{-16}$ is presumably caused by a piece of about 20~m yet unstabilized fiber between a cw laser and one of the frequency combs. Thus, at this time scale the measurement can be considered as a conservative estimate of relative noise between two independent multi-port comb systems. In future, the unstabilized fiber will be included in the active noise cancellation to provide a fully symmetric setup and we are confident to substantially reduce the instability plateau to below $1\times10^{-16}$.

The inter-branch performance does not seem to be affected by fiber noise. The reason becomes clear when looking at the branch layouts. Both conversion branches (VIS and IR) share the same environmental shielding so that most of the residual fiber noise is common mode. In conclusion, the inter-comb performance can be further improved employing temperature stabilisation or acoustic shielding to the entire fiber comb setups.

On very long time scales ($\approx10^4$~s), the inter- and inner-branch measurement seem to level out in the $10^{-19}$ instability range. The reason for this is not fully understood. A possible explanation is the temperature dependence of the employed rf electronics that sets in and limits the measurements. However, this is about two orders of magnitude lower than what is currently achieved with the best clocks. Also the inter-comb experiment averages down to $3\times10^{-18}$, still surpassing the best clock comparisons by a factor of three \cite{ros08,fal11,hun12}.

\section{Spectroscopy and clock operation with $^{87}$Sr}
\label{sec:sr87}

\begin{figure}[!t]
\centering
\includegraphics[width=\columnwidth]{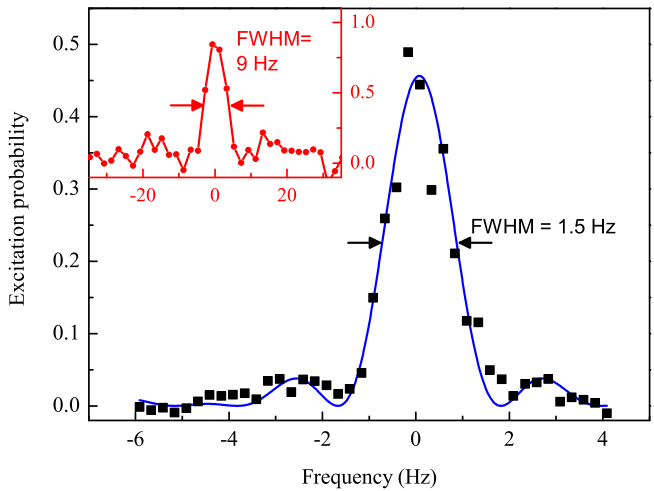}
\caption{Spectra of one Zeeman component of the $^{87}$Sr clock transition with the strontium clock laser only (inset, 10~Hz line width, data from Ref.~\cite{fal11}) and with the stability transfer from the master laser stabilized to the cryogenic silicon cavity.}
\label{fig:linescan}	
\end{figure}

To demonstrate the feasibility and usefulness of stability transfer from a extremely high quality oscillator to the interrogation laser of an optical clock, we have used the PTB strontium lattice clock \cite{fal11} as test system. The lattice clock setup and its clock laser have been described in some detail in previous publications \cite{leg09,lis09,vog11,mid11}. Thus, we focus here on the central aspects of the experiment only. 

The strontium clock laser is stabilized to a reference resonator with a spacer made from ULE glass to which mirrors with ULE substrates are optically contacted. The laser provides a fractional short-term stability of $\sigma_y\approx2\times10^{-15}$ for averaging times of a few seconds.
Using this laser we typically observe spectra of the $^1$S$_0 - ^3$P$_0$ clock transition at 698~nm in $^{87}$Sr with a Fourier limited line width of 10~Hz and 90~\% contrast (inset in Fig.~\ref{fig:linescan}).

To record the spectra the atoms are loaded from a Zeeman-slowed optical beam into a magneto-optical trap (MOT) operated on the 462~nm resonance line of strontium. The temperature of the atomic cloud is reduced to a few micro-Kelvin in a second MOT phase employing the intercombination line $^1$S$_0 - ^3$P$_1$ (689~nm; 7~kHz line width). This temperature is low enough to load the atoms into an optical lattice operated at the light-shift cancellation wavelength for the clock transition.

\begin{figure}[!t]
\centering
\includegraphics[width=\columnwidth]{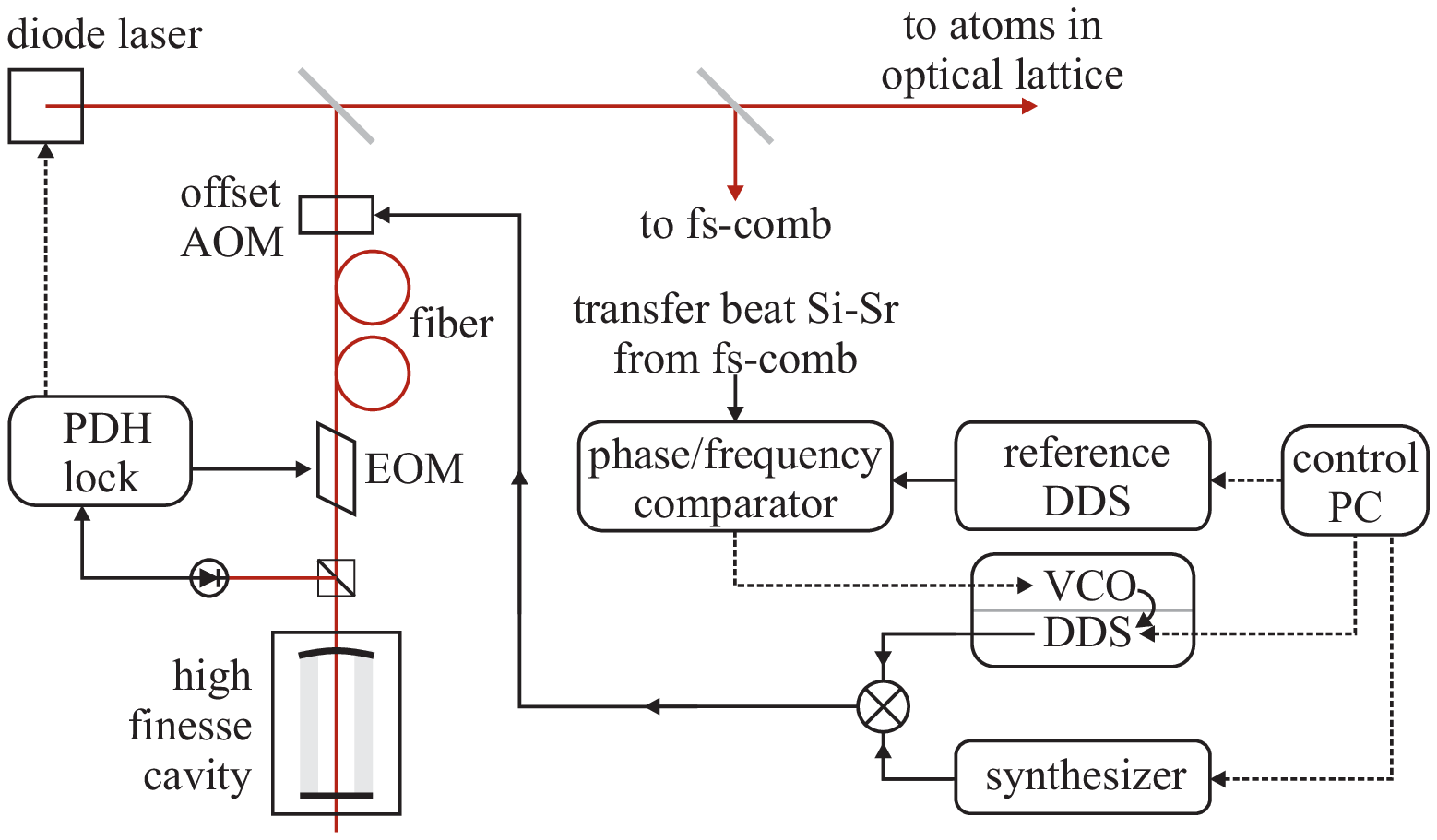}
\caption{Schematics of the lock of the Sr clock laser to its reference resonator (left; PDH: Pound-Drever-Hall; AOM: acousto-optical modulator; EOM: electro-optical modulator) providing a short-term stability of $\sigma_y\approx2\times10^{-15}$ at a few seconds. The rf driving the offset AOM is generated by mixing of two computer controlled sources. To phase lock the clock laser to the master laser, an error signal is generated in a phase/frequency comparator, to which the transfer beat from the fs frequency comb and a stable rf are fed.}
\label{fig:phaselocksr}	
\end{figure}

To lock the laser frequency to the strontium clock transition we spin-polarize the atoms in the optical lattice to either one of the extreme Zeeman levels. They are probed in a Rabi-type interrogation on the low and high frequency half width points of the lines in a weak magnetic field to separate the Zeeman transitions. For clock operation the frequency of the clock laser is steered to the average of the two transition frequencies using an offset AOM (Fig.~\ref{fig:phaselocksr}), which cancels the linear Zeeman shift.

Previously, we have estimated the clock stability by interleaving two stabilizations and calculating the Allan deviation of the difference of the two steering frequencies applied to the offset AOM \cite{fal11}. We found that the instability of this difference averages as $5\times 10^{-15}/\sqrt{\tau/ {\rm s}}$, meaning that this instability is available for the evaluation of e.g. systematic effects. In non-interleaved clock operation the instability is expected to be a factor of at least two smaller because the cycle time, the two-fold noise contribution, and the Dick effect \cite{que03} are reduced -- provided the transition frequency does not show time dependent shifts.

To make use of the superior stability of the master laser (see Section~\ref{sec:master}), i.e. to improve the stability of the lattice clock, we have phase locked the Sr clock laser (slave laser) with a bandwidth of about 500~Hz to the master laser using a femtosecond comb as transfer oscillator (see Section~\ref{sec:transfer}). Due to the small bandwidth the contribution of the white phase noise is suppressed and the performance is expected to be limited by acoustic and thermal influences on the fiber. The error signal derived from phase comparison of transfer beat \cite{leg09} with a stable rf reference is used to steer the radio frequency driving the offset AOM between clock laser and its ULE reference cavity (Fig.~\ref{fig:phaselocksr}). This radio frequency is the sum of two oscillators \cite{fal12} that independently provide a drift compensation for the ULE cavity using a direct digital synthesis (DDS) frequency generator and stepwise corrections to lock the laser to the clock transition via a conventional frequency synthesizer. The error signal of the phase lock acts on the VCO serving as local oscillator in the DDS generator. To scan the clock laser over or to lock it to the clock transition, the reference DDS of the phase lock is adjusted in addition to the conventional frequency synthesizer by the data acquisition system. The adjustment of the synthesizers frequency serves as feed forward to minimize the necessary adjustments of the drift compensating DDS.

\begin{figure}[!t]
\centering
\includegraphics[width=\columnwidth]{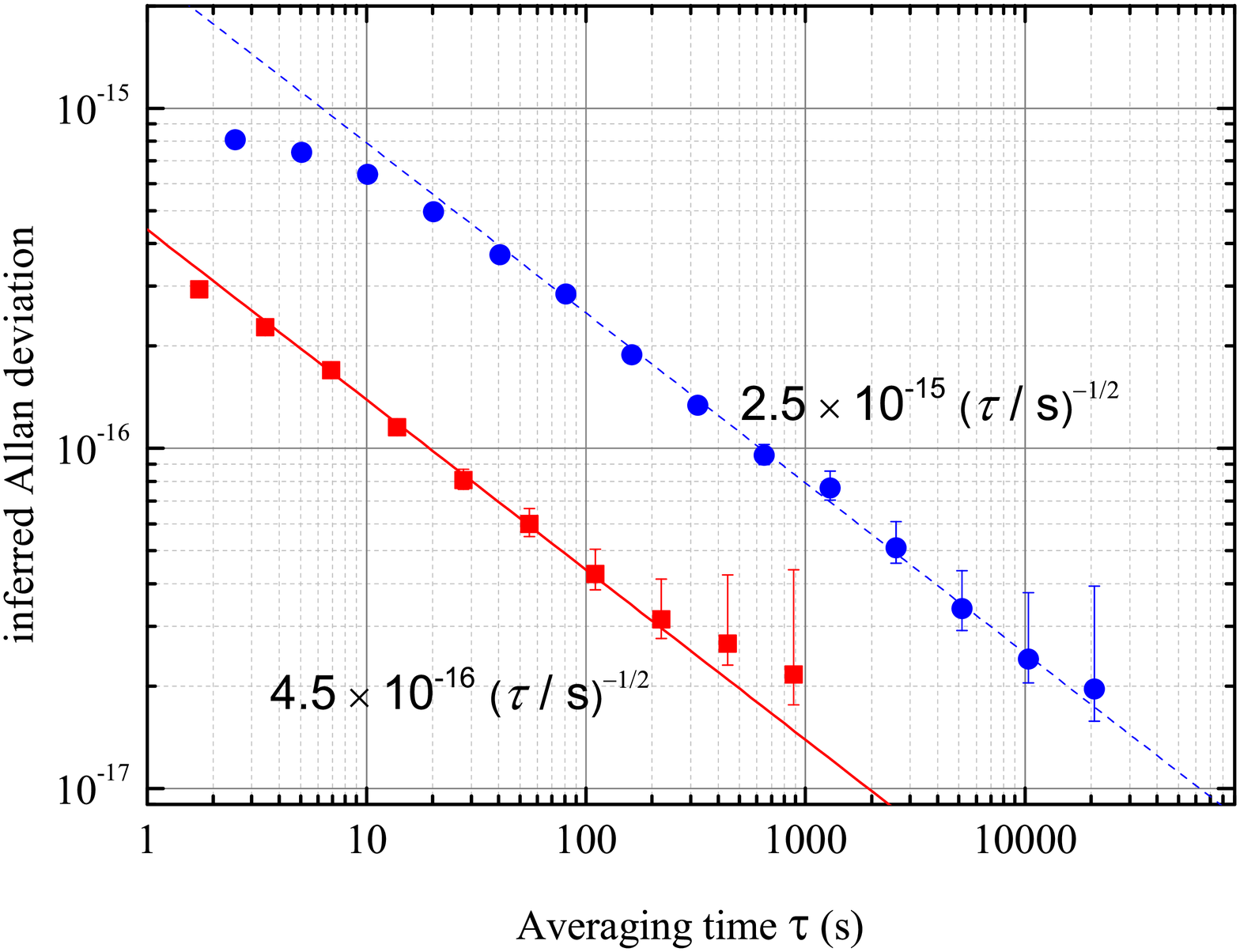}
\caption{Stabilities of the Sr lattice clock inferred from interleaved measurements using the stability of the ULE reference cavity of the Sr clock laser (upper curve, data from Ref.~\cite{fal11}) and with the stability transfer from the silicon cavity stabilized laser system (lower curve). Averaging times to achieve a desired statistical uncertainty are reduced in the latter case by a factor of 30.}
\label{fig:interleaved}	
\end{figure}

Activating the phase lock we have recorded spectra of the clock transitions with line width reduced to below 2~Hz without loss of contrast (Fig.~\ref{fig:linescan}). Longer interrogation times at reduced laser intensity did not provide narrower transitions and caused a considerable reduction of excitation probability. We did not observe an excitation probability close to unity for any interrogation pulse length since we employed an interrogation sequence that did not remove atoms remaining in other than the interrogated Zeeman levels after imperfect spin polarization.

To estimate the clock stability, we simply locked the laser frequency to a single Zeeman component, interleaving to stabilization cycles. This is sufficient to remove instabilities in the interleaved signal originating from e.g. drifts of the magnetic field and shortens the servo time. A single interrogation on one half width point of a Zeeman component including the preparation and detection of atoms required 860~ms with a duty cycle of 55~\%. We observe a significantly reduced instability (Fig.~\ref{fig:interleaved}): We infer a non-interleaved clock stability of $4.5\times10^{-16}/ \sqrt{\tau/ {\rm s}}$ that is similar to the best so far achieved values \cite{tak11,jia11}. This improvement of clock stability reflects mostly the improvement of the line quality factor by six (10~Hz/1.5~Hz). The cycle time $T_{\rm C}$ was not considerably increased in the current experiment (860~ms) compared to the previous ones (630~ms) since the loading and preparation time could be reduced significantly. As the clock stability scales with $\sqrt{T_{\rm C}}$ (neglecting the Dick effect), it is only slightly degraded by the increase of cycle time. In conclusion it is obvious that the improved clock stability facilitates the clock operation enormously as averaging times for many investigations are reduced by a factor of~30.

\section{Conclusion and Outlook}
\label{sec:outlook} 
Bridging the spectral gap with a femtosecond fiber comb, we transferred the 10$^{-16}$-level short-term stability of a high-end cryogenic cavity stabilized laser (master laser) operated at 1.5~$\mu$m to a clock laser (slave laser) operated at 698~nm. We have investigated potential noise sources arising from independent signal processing in the different branches of the femtosecond fiber comb. Our measurements have shown that the short-term performance of the transfer is in a range of $\sigma_y\approx1\ldots3\times10^{-16}$ at a second, which is already on the level of stability of the best current lasers. This is however not a principle limitation, as environmental perturbations can be reduced by temperature stabilisation and acoustic shielding of the fiber comb setup.

Despite the current limitation of the setup, the stability of the slave laser was significantly improved -- resolving the $^1$S$_0 - ^3$P$_0$ clock transition of $^{87}$Sr to a Fourier-limited line width of 1.5~Hz. Furthermore, we expect clock instabilities to average down with $4.5\times10^{-16}/ \sqrt{\tau/ {\rm s}}$ as inferred from an interleaved stabilisation. This compares with best so far observed clock instabilities \cite{jia11}. 

Enabling stability transfers across the full wavelength range covered by the frequency comb, this concept allows to focus efforts in the development of a single outstanding oscillator and provide its stability for simultaneous operation of multiple clocks.

\section{ACKNOWLEDGMENT}
We thank B. Lipphardt for his contributions to the inter-comb experiment. The cryogenic silicon cavity laser system employed in this work is developed
jointly by the JILA Physics Frontier
Center (NSF) and the National Institute of Standards
and Technology (NIST), the Centre for Quantum Engineering and
Space-Time Research (QUEST) and Physikalisch-Technische
Bundesanstalt (PTB). Financial support from the European Communitys ERA-NET-Plus Programme (Grant No. 217257) and from the European Communitys Seventh Framework Programme (Grant Nos. 263500 and JRP IND14) is gratefully acknowledged.

\end{document}